\documentclass[9pt,twocolumn,twoside]{optica}
\setboolean{shortarticle}{true}
\setboolean{minireview}{false}
\title{Wide-field SU(1,1) interferometer}

\author[1,2,*]{G.~Frascella}
\author[3]{E.~E.~Mikhailov}
\author[4]{N.~Takanashi}
\author[5,6]{R.~V.~Zakharov}
\author[5,6]{O.~V.~Tikhonova}
\author[1,2,5]{M.~V.~Chekhova}

\affil[1]{Max-Planck Institute for the Science of Light, Staudtstr. 2, Erlangen D-91058, Germany}
\affil[2]{University of Erlangen-Nuremberg, Staudtstr. 7/B2, 91058 Erlangen, Germany}
\affil[3]{Department of Physics, College of William \& Mary, Williamsburg, Virginia 23187, USA}
\affil[4]{Department of Applied Physics, School of Engineering, The University of Tokyo, 7-3-1 Hongo, Bunkyo-ku, Tokyo 113-8656, Japan}
\affil[5]{Physics Department, Moscow State University, Leninskiye Gory 1-2, Moscow 119991, Russia}
\affil[6]{Skobeltsyn Institute of Nuclear Physics, Lomonosov Moscow State University, Moscow 119234, Russia}
\affil[*]{Corresponding author: gaetano.frascella@mpl.mpg.de}

\begin{abstract}
An ${\rm SU(1,1)}$ interferometer uses a sequence of two optical parametric
amplifiers for achieving sub-shot-noise sensitivity to a phase shift introduced in between.
We present the first realization of a wide-field ${\rm SU(1,1)}$
interferometer, where the use of a focusing element enables spatially multimode operation within
a broad angle. Over this angle, the interference phase is found to be flat. This property is important for the high sensitivity to the phase front disturbance. Further, $-4.3\pm0.7$ dB quadrature squeezing, an essential requirement to the high sensitivity,
is experimentally demonstrated for plane-wave modes inside the interferometer. Such an interferometer is useful not only for
quantum metrology, but also in remote sensing, enhanced sub-shot-noise imaging,
and quantum information processing.
\end{abstract}

\setboolean{displaycopyright}{true}

\begin{document}

\maketitle

Interferometers have been used for more than a century to measure
physical quantities with high accuracy. Recently,
the experimental realization of ${\rm SU(1,1)}$ interferometers has
raised significant interest due to the loss-tolerant sub-shot-noise sensitivity~\cite{Li:14,Hudelist:14,Chekhova:16,Lemieux:16,Linneman:16,Lukens:16,Anderson:17,Gong:17,Manceau:17PRL,Du:18,Liu:18,Lukens:18,Liu:18fibers,Shaked:18}.
The core idea is to use a series of two optical parametric amplifiers
(OPAs) to probe phase shifts between them~\cite{Yurke:86,Sparaciari:16}. Possible
applications are in remote sensing~\cite{Liu:18} and in
quantum information processing~\cite{Shaked:18}, but such a scheme
is especially attractive for quantum metrology with optical
\cite{Hudelist:14,Anderson:17,Manceau:17PRL}, atom~\cite{Gross:10,Linneman:16}
and hybrid~\cite{Chen:15} interferometers. 

The two-mode squeezed state employed in an ${\rm SU(1,1)}$
interferometer is a quantum resource that helps to overcome the limitations
encountered in a classical framework. In the optical
phase measurement, the achievable sensitivity for an ${\rm SU(1,1)}$ interferometer
beats the shot-noise limit, especially in the low photon-number
regime~\cite{Manceau:17PRL,Anderson:17}. The $\mathrm{SU(1,1)}$
configuration is tolerant to detection losses, provided that the second
OPA has a sufficiently large gain~\cite{Manceau:17,Giese:17,Shaked:18}.
To satisfy this condition, one can use two nonlinear $\chi^{\left(2\right)}$
crystals as OPAs, since unbalancing the OPAs does
not highly affect the mode composition, compared to the case of atoms
\cite{Gross:10,Linneman:16} or four-wave mixers~\cite{Hudelist:14,Lukens:16}. 
All ${\rm SU(1,1)}$ interferometers realized so far are spatially
single-mode and allow high sensitivity in one dimension only. 

In this Letter, we report the first demonstration of a spatially multimode
${\rm SU(1,1)}$ interferometer, using high-gain parametric down-conversion
(PDC) produced in a nonlinear crystal. This paves the way towards
2D phase profiling in the quantum regime, close to the common idea
of an interferometer with a 2D fringe pattern. Furthermore, our configuration
opens up the possibility to enhanced quantum imaging~\cite{Knyazev:19} and to the detection of a small disturbance
in the orbital angular momentum (OAM)~\cite{Liu:18}. Finally,
we report the measurement of quadrature squeezing for plane-wave modes inside the interferometer,
as a prerequisite for achieving high sensitivity. We find that for all plane-wave modes,
it is approximately the same quadrature that is squeezed.

\begin{figure}[htb]
\begin{centering}
\includegraphics[width=0.37\textwidth]{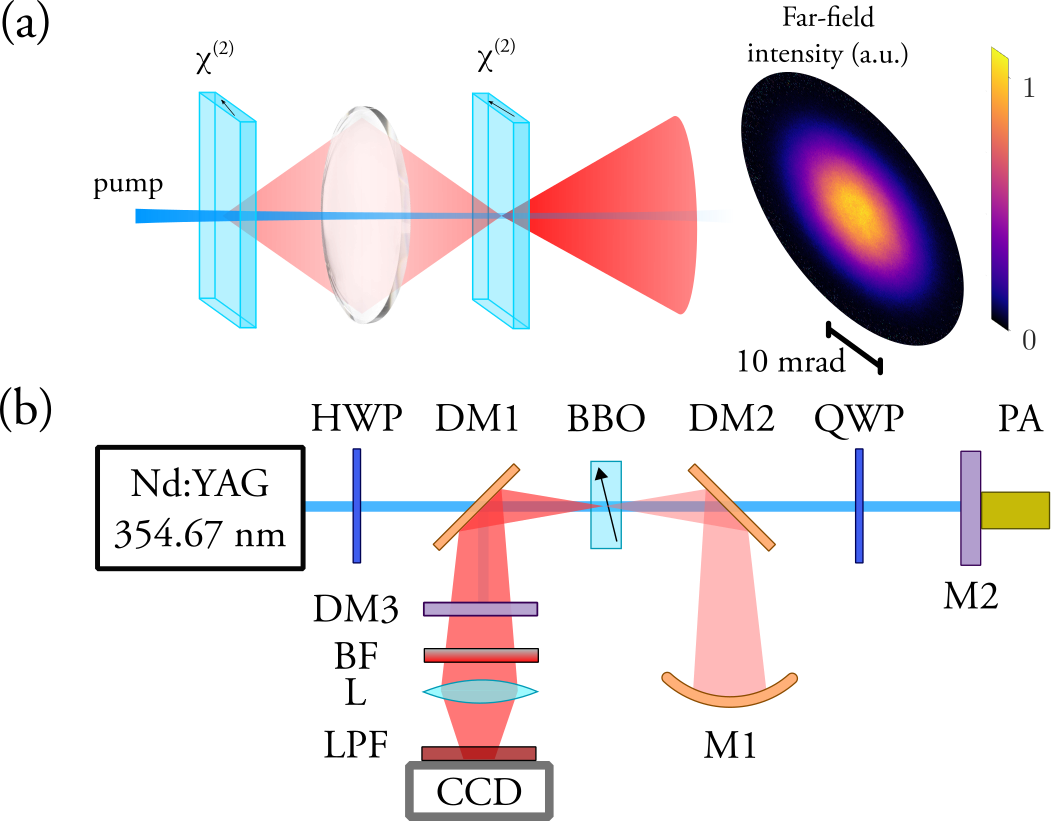} 
\par\end{centering}
\caption{(a) In a wide-field ${\rm SU(1,1)}$ interferometer, the 
degenerate PDC radiation from the first $\chi^{\left(2\right)}$ crystal
is imaged into the second one. This enables multimode amplification/de-amplification, shown in the experimental far-field
image. (b) In our experiment, the pump and the PDC are split with dichroic mirrors DM1-2.
A spherical mirror M1 images the PDC radiation
onto the crystal. Mirror M2 is mounted on the piezoelectric actuator
PA to control the phase and the half-wave plate HWP and quarter-wave
plate QWP control the polarisation. The amplified/de-amplified PDC radiation is observed in the
Fourier plane of lens L on the CCD camera. Dichroic mirror
DM3 rejects the pump, while bandpass filter BF and long
pass filter LPF provide spectral filtering.}
\label{fig:setup} 
\end{figure}
The principle of our $\mathrm{SU(1,1)}$ interferometer is based on two nonlinear crystals with a focusing
element in between, as shown in Fig. \ref{fig:setup} (a). The PDC radiation produced in
the first crystal in the degenerate type-I regime (shown with a red filled cone) is amplified or de-amplified
in the second crystal (shown as a brighter cone) depending
on the optical phase between the pump, signal and idler fields $\phi=\phi_{p}-\phi_{s}-\phi_{i}$~\cite{Manceau:17}.
 To exploit the full multimode structure of the
radiation generated in the first crystal, the divergence of the PDC
light is corrected with a lens. Provided that the PDC generation
region in the first crystal is imaged into the second crystal, the
amplification occurs for all angles of emission, intrinsically restricted
only by the phase matching conditions. The configuration presented
here offers good visibility over broad angles due to the mode matching. In previous experiments without the focusing element,
spatial filtering of a single spatial mode around the pump direction
was necessary to obtain high visibility and achieve sub-shot-noise phase sensitivity~\cite{Manceau:17PRL}. 

To avoid the focusing of the pump beam, the
paths of the pump and the PDC radiation are split and folded,
see Fig. \ref{fig:setup} (b). The pump is the third harmonic
of a pulsed ${\rm Nd:YAG}$ laser (wavelength $354.67$ nm, repetition
rate $1$ kHz, pulse duration $18$ ps, average power 60 mW). Type-I collinear degenerate PDC is generated in a $\beta$-barium borate (BBO) crystal. The
half-wave plate HWP misaligns the linear polarization of the pump
by $27$ deg with respect to the optimal (horizontal) direction in order to reduce the parametric gain in the first pass. 

Through the dichroic mirror DM2, the PDC radiation is sent to the
focusing element, i. e. a spherical mirror M1 with curvature radius
$R=100$ mm, and is imaged back onto the crystal. The pump transmitted
by DM2 is sent to the quarter-wave plate QWP for polarization control
and to the mirror M2 mounted on a piezoelectric actuator PA for phase
control. The QWP on a double pass acts as an HWP and,
by rotating the pump polarization, controls the
parametric gain of the PDC generated in the second pass. Additionally, the pump has a beam size of FWHM $300\pm10$
$\mu{\rm m}$ in the first pass and, to provide a higher parametric
gain, $180\pm10$ $\mu{\rm m}$ in the second pass. From the nonlinear dependence of the PDC intensity
on the pump power, $I(P)\propto\sinh^2G,\,G\propto\sqrt{P}$, we measure the gain $G$~\cite{Iskhakov:12}. We obtain separately the
gain from the first pass $G_{1}=2.1\pm0.3$ and from the second pass
$G_{2}=3.3\pm0.3$. The gain unbalancing does not reduce much the interference visibility (see \href{link}{Supplement 1} for details)
but it is crucial for the detection of the squeezing, as explained below.

If the path lengths from M1 and M2 to the crystal
are such that the pump and PDC radiation pulses overlap on the way back, phase-sensitive amplification/de-amplification
occurs depending on the phase shift controlled by the PA. The phase can be locked with the use of an additional
beam injected at the unused port of DM2 and a feedback circuit (not
shown). The dichroic mirror DM3 reflects the amplified/de-amplified PDC radiation to a charge-coupled
device (CCD) camera in the Fourier plane of lens L with a focal
length $f=40$ mm. The spectral filtering is achieved with a bandpass
filter BF (central wavelength either $700$ or $710$ nm, bandwidth
$10$ nm) and a long-pass filter LPF with the edge at $645$ nm, directly
attached to the CCD.

\begin{figure}[htb]
\begin{centering}
\includegraphics[width=0.44\textwidth]{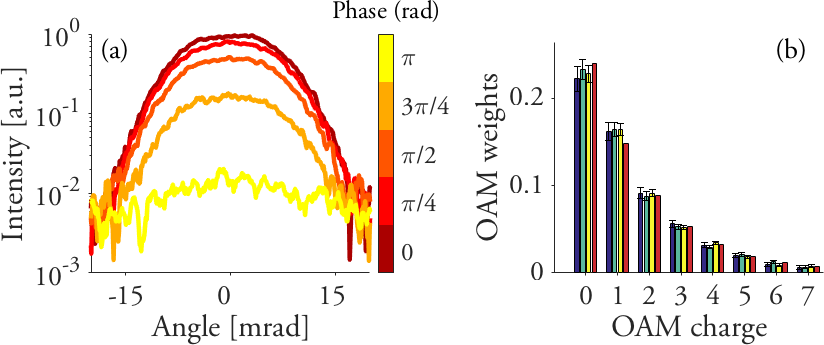} 
\par\end{centering}
\caption{(a) Intensity profile at the output of the interferometer
for different phases. (b) Measured weights of OAM modes at the output of the interferometer show independence on the phase as predicted by theory (red). Relative shifts
of the interferometric phase with respect to the dark fringe $\phi=\pi$
are $+0.68$ rad (blue), $+0.88$ rad (green), $+1.08$ rad (yellow).}
\label{fig:OAM} 
\end{figure}
The profile of the single-shot intensity distribution measured at
the output is shown in Fig. \ref{fig:OAM} (a). As the phase is scanned,
the amplitude of the characteristic flat-top intensity distribution
for phase-matched PDC emission varies periodically, while the width
stays constant. As shown in Fig. \ref{fig:OAM} (a), the visibility
of the interference pattern for a single pixel is $98\%$, but even
for the intensity integrated in two dimensions it is more than $95\%$. 

Another remarkable property of the interferometer presented here is
the stability of the spatial spectrum at the output, and, in particular,
of the OAM spectrum, as the phase changes. Indeed, for any kind of
application, the operation of the interferometer needs to be the same
for all spatial modes. Further, the change of the OAM spectrum due
to an azimuthal phase perturbation can be sensitively detected.
To obtain the OAM spectrum at the output, we use a method based on the measurement of the covariance of intensities at different far-field points~\cite{Beltran:17}, specified by the transverse wave-vectors $\vec{q},\vec{q'}$, with modules $q,q'$ and azimuthal angles $\theta,\theta'$. 
Under the condition $q=q'=q_{0}$, the covariance, formally defined as 
$\mathrm{Cov}_{\left|q=q'=q_{0}\right.}\left(\theta,\theta'\right)=\left<I\left(\theta\right)I\left(\theta'\right)\right>-\left<I\left(\theta\right)\right>\left<I\left(\theta'\right)\right>$,
can be shown to depend only on the difference $\Delta\theta=\theta-\theta'$~\cite{Beltran:17}. Moreover, if the cross-correlations in the PDC radiation are removed by filtering a slightly non-degenerate wavelength, the covariance averaged over the variable $q_0$ is 
\begin{equation}
\mathrm{Cov}\left(\Delta\theta\right)=\left[\sum_{l=-\infty}^{\infty}\Lambda_{l}e^{il\Delta\theta}\right]^{2}.\label{eq:azimcov}
\end{equation}
Here, $\Lambda_{l}$ are the weights of modes with OAM $l$, normalized as $\sum_{l=-\infty}^{\infty}\Lambda_{l}=1$. They can be extracted
by performing a Fourier decomposition on the square root of Eq. \ref{eq:azimcov}. In experiment, we removed the cross-correlations with a $10$ nm bandpass filter centered at $700$ nm, which is shifted with respect to the degenerate frequency $709.33$ nm. The covariance was measured from $\sim500$ single-pulse intensity spectra acquired with the CCD camera.

Figure~\ref{fig:OAM} (b) shows the OAM weights of the radiation at
the output of the interferometer for three different optical phases,
namely $+0.68$ rad, $+0.88$ rad and $+1.08$ rad from the dark fringe
respectively with blue, green and yellow bars. The weights remain
the same as the phase is changed and this demonstrates
that all OAM modes are uniformly amplified/de-amplified. The negative-OAM part
is not shown here since
it is symmetric with respect to $l=0$. 
The effective number of OAM modes, given by the
formula $\left(\sum_{l=-\infty}^{\infty}\Lambda_{l}^{2}\right)^{-1}$,
is $7.6\pm0.2$. For comparison, the number of {OAM modes for}
the radiation in the first pass is theoretically estimated to be $13$.
The double-pass configuration reduces the number of modes since the
effective gain is larger than the gain of the first pass~\cite{Sharapova:15}. 
We obtain the theoretical distribution of the OAM weights through the
Schmidt decomposition of the two-photon amplitude of PDC for the double-pass configuration
\cite{Sharapova:15,Zakharov:18}. The theory (red
bars in Fig. \ref{fig:OAM} (b)) confirms the experimental results in the same range of phases. For the case $\phi=\pi$, the theoretical distribution shows a slower decay and a $18\%$ increase
in the effective number of modes.

Estimating the number of radial modes at the output of the interferometer
to be $4.5\pm0.2$ from a similar experimental reconstruction in the
radial degree of freedom, the total number of spatial modes is $35\pm3$. 
The multimode feature is attractive for the realization of high-dimensional
quantum spaces~\cite{vandeNes:06,Molina:07}, but also for imaging
\cite{Boyer:08Scie}. Indeed, the high-order modes are connected with
fine details, because of their high spatial frequency, and represent
a resource for the resolution in imaging experiments. We therefore characterize our interferometer as a `wide-field' one: it provides both a relatively broad angle ($20$ mrad) and a large number of angular modes within this range.

Finally, we show that the quadrature noise 
for the radiation inside the interferometer is below the shot noise.
This is fundamental to achieve enhanced sensitivity with respect to a 
`classical' interferometer with the same number of photons. 
Homodyne detection is not suitable in our case since the PDC emission is highly multimode and it requires
the appropriate shaping of the local oscillator field for the measurement
of the squeezing in particular eigenmodes (one at a time)~\cite{Bennink:02, Embrey:15}.
However, it was recently shown~\cite{Shaked:18} that the second amplifier in an ${\rm SU(1,1)}$ interferometer can be used for an `optical homodyne' measurement of quadrature squeezing at the output of the first amplifier. In this approach, the variances of the input quadratures $\hat{x}_{i},\hat{p}_{i}$ can be found
by measuring the total intensity at the output:
\begin{equation}
I=C\cdot{\rm Var}\left(\hat{x}_{\psi}\right),
\label{eq:parametric-amplif}
\end{equation}
with the calibration constant $C$ and the generic quadrature operator
$\hat{x}_{\psi}=\hat{x}_{i}\cos\psi+\hat{p}_{i}\sin\psi$ and the phase of the interferometer $\phi=2\psi$. Eq. \ref{eq:parametric-amplif}
is valid only under the assumption $e^{4G_{2}}{\rm Var}\left(\hat{x}_{i}\right)\gg {\rm Var}\left(\hat{p}_{i}\right)$, where 
$\hat{x}_{i}$ is the squeezed quadrature. This holds true in our case because of the unbalanced gains and because ${\rm Var}\left(\hat{p}_{i}\right)/{\rm Var}\left(\hat{x}_{i}\right)$ cannot exceed $e^{4G_{1}}$. 
The constant $C$ can be calibrated
by removing the input to the second-pass OPA, leaving only vacuum
fluctuations. Any loss induced at the detection stage will not contribute 
because it is already included in the constant $C$.
This consideration is valid for single mode, but it can be
generalized to the multimode scenario. In the measurement
of the total intensity at the output, the contribution of each mode to the amplification/de-amplification
depends on the relative phase between the modes and on the overlap
of the modes of the input state with the modes generated in the second-pass
OPA~\cite{Wasilewski:06}. The shapes of the modes for PDC radiation
change very little as the gain increases~\cite{Sharapova:18}, therefore,
the overlap can be reasonably high.

\begin{figure}[htb]
\begin{centering}
\includegraphics[width=0.51\textwidth]{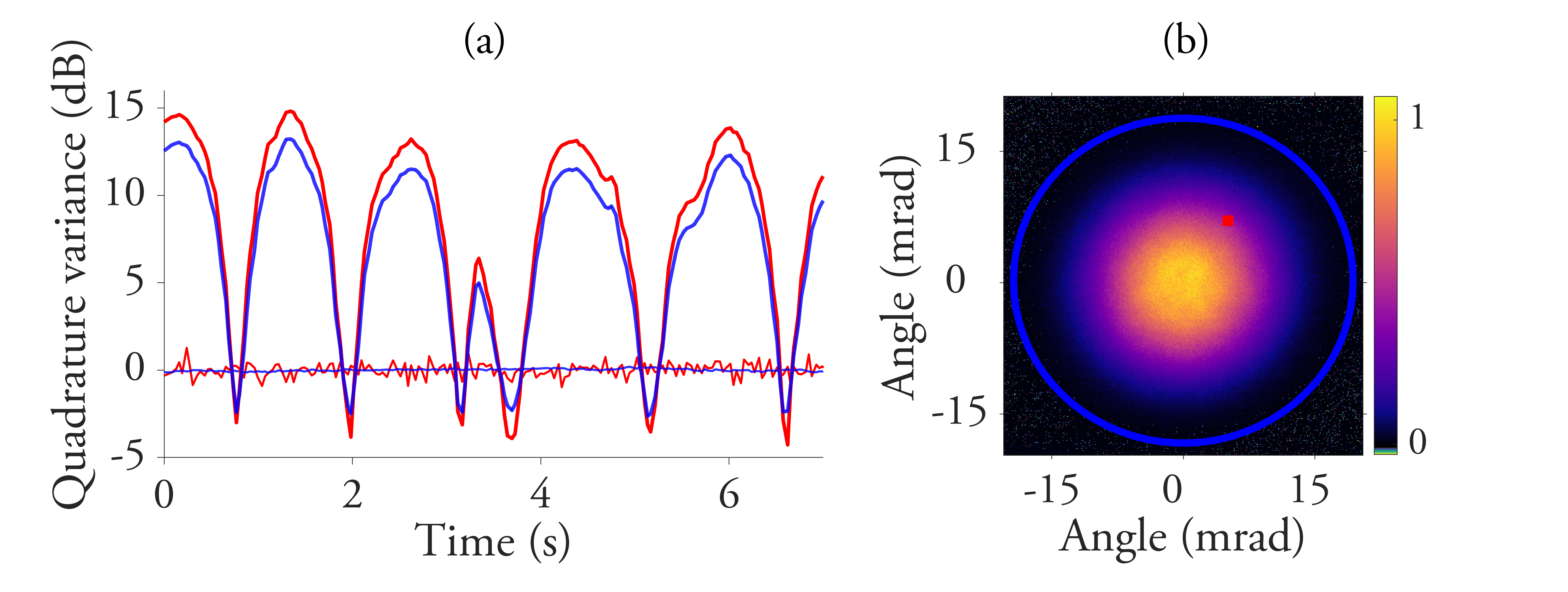} 
\par\end{centering}
\caption{(a) Quadrature variance in dB changes with time for a triangle-wave
scan of the piezo actuator. The traces around zero show the shot-noise level. The two different
colors correspond to the measurement on the full frame (blue) and a
single pixel (red), as shown on the intensity distribution at the
output of the interferometer in panel (b).}
\label{fig:squeezing} 
\end{figure}
The level of amplification/de-amplification is measured by scanning
the phase with a triangle-wave voltage applied to the PA and the result
is shown in Fig.~\ref{fig:squeezing} (a). The measurement can be
considered simultaneously for two different regions
of interest (ROI) shown in panel (b): a single pixel
(red) and the total frame (blue). The constant $C$ of Eq. \ref{eq:parametric-amplif}
is obtained by blocking the radiation from the first pass at the curved
mirror M1 (fluctuations are shown with the traces around zero). The
small and large ROI give respectively the best squeezing level of
$-4.3\pm0.7$ dB, $-2.6\pm0.3$ dB and the best anti-squeezing level
of $14.8\pm0.9$ dB, $13.2\pm0.1$ dB. There is a good agreement between
the anti-squeezing level (less sensitive to mode mismatch and losses)
and the value expected from the independently measured gain $G_{1}$.
The measurement is made with the bandpass filter centered at $710$ nm, but the result is similar for $700$ nm.

\begin{figure}[htb]
\begin{centering}
\includegraphics[width=0.45\textwidth]{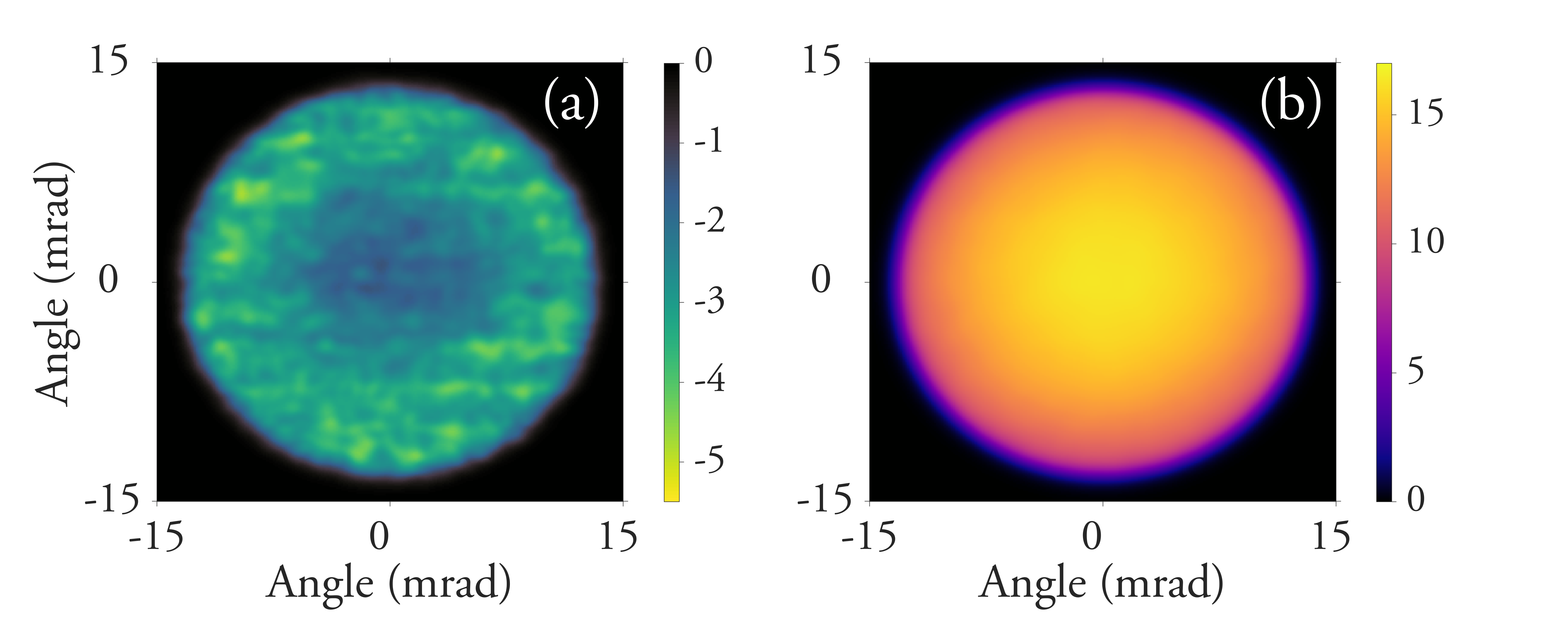} 
\par\end{centering}
\begin{centering}
\includegraphics[width=0.45\textwidth]{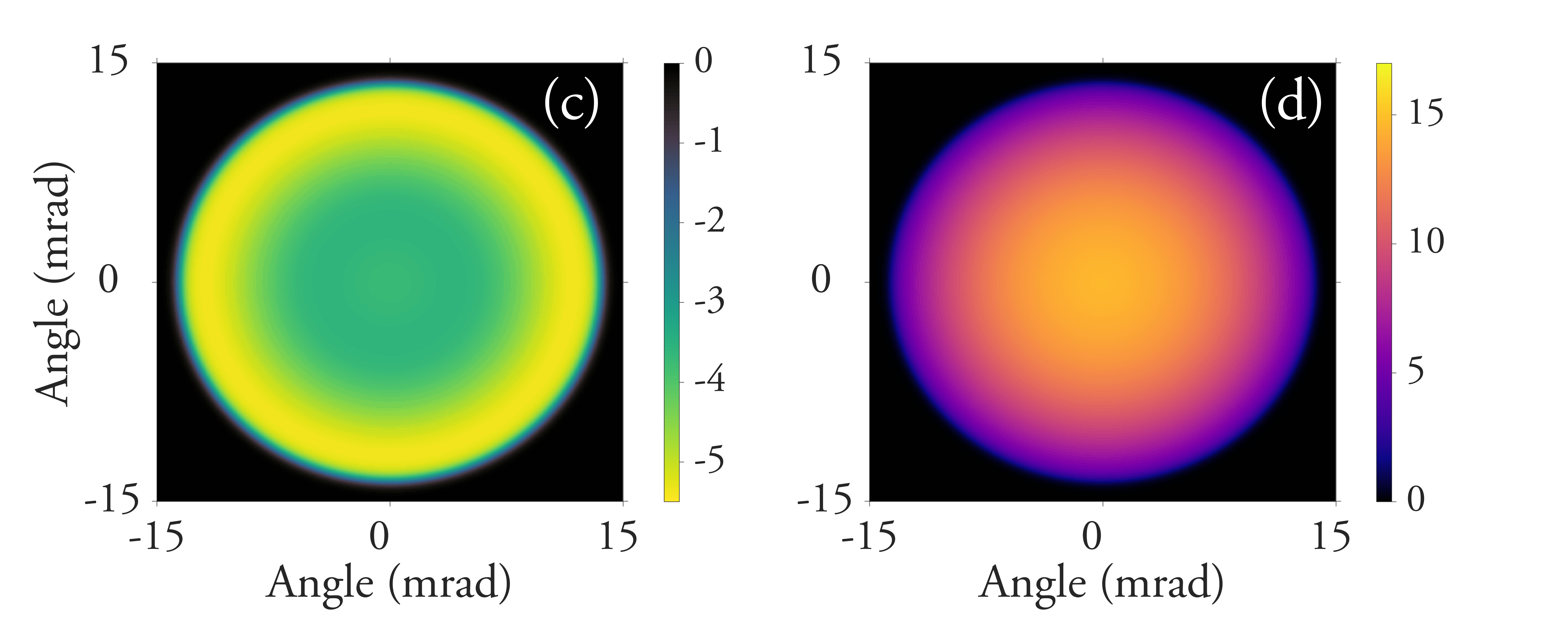}
\par\end{centering}
\caption{`Optical homodyne' measurement of quadrature
squeezing (a) and anti-squeezing (b) for different plane-wave modes and its theoretical simulation (c,d).}
\label{fig:squeezing-map} 
\end{figure}
By considering every pixel of the CCD, we measure the quadrature variance for all plane-wave modes, as shown in Fig. \ref{fig:squeezing-map}.
Panel (a) is the 2D distribution of the squeezing, while panel
(b) shows the anti-squeezing. The sharp borders to zero are applied in both panels to remove artifacts caused by very low intensity. One should
expect the distributions in Fig. \ref{fig:squeezing-map} to be flat,
but in our case, the slight difference in the phase-matching of
PDC in the first and second passes leads to squeezing slightly 
changing towards the center. Indeed, the intensity distribution
from the second pass is $35\%$ broader than the one from the first
pass, because a slight misalignment of the pump beam at the mirror
M2 modifies the phase-matching conditions. This effect leads to mode mismatch, which affects mainly the squeezing distribution. The theoretical simulation of this behaviour, shown in Fig. \ref{fig:squeezing-map}
(c) and (d) respectively for squeezing and anti-squeezing, is in good
agreement with the experiment.

The wide-field ${\rm SU(1,1)}$ interferometer can be expanded to place 
inside the interferometer an object, whose absorption 
or image can be detected with enhanced sub-shot-noise accuracy 
and tolerance to detection losses~\cite{Knyazev:19}. 
%In addition, if the movement of an object slightly disturbs the OAM spectrum inside the interferometer, 
%it should be possible to sense the disturbance remotely~\cite{Liu:18}. 
Further, the scheme
presented here can be used in information processing
as it provides a `quantum network', i.e. a multimode (multipartite) quantum system,
similarly to other realizations in space~\cite{Armstrong:12} and
frequency~\cite{Cai:17}. Indeed, the multipartite entanglement
depends on the allocation of the spatial modes specified by the user
and should be readily available with the selection in the detection
process. The important advantage introduced here is that detection losses do not contribute to the measurement
of the quadrature noise. In our case, one could use a particular mode
combination without worrying about the effect of losses and simulate
a linear optical network with the change of basis. 

In conclusion, we have constructed a spatially multimode ${\rm SU(1,1)}$
interferometer by introducing a focusing component between the two
OPAs. We have proved that the interferometer has the same multimode
structure (around $35$ modes) as the phase changes. The quadrature
noise for the radiation inside the interferometer has been proved
to be below the shot noise (with the best squeezing level of $-4.3$
dB), as required for the highly sensitive detection of phase shifts.
The measurement of the 2D distribution of the amplification/de-amplification
level for the plane-wave modes reveals the uniform amplification phase
for the quadrature variance of such modes. Possible applications
of such an interferometer are the detection of the disturbance in
the OAM imparted by an object, imaging with
sub-shot-noise precision, quantum information processing in a
multidimensional space and quantum metrology in two dimensions.

\section*{Funding Information}
We acknowledge financial support of the Russian Science Foundation Project No.19-42-04105 and of the Japan Society for the Promotion of Science, Overseas Challenge Program for Young Researchers.

\bigskip \noindent See \href{link}{Supplement 1} for supporting content.

% Bibliography

\bibliography{widefieldSu11}

\begin{thebibliography}{10}
\newcommand{\enquote}[1]{``#1''}

\bibitem{Li:14}
D.~Li, C.-H. Yuan, Z.~Y. Ou, and W.~Zhang, {\protect\JournalTitle{New Journal
  of Physics}} \textbf{16}, 073020 (2014).

\bibitem{Hudelist:14}
F.~Hudelist, J.~Kong, C.~Liu, J.~Jing, Z.~Y. Ou, and W.~Zhang,
  {\protect\JournalTitle{Nature Communications}} \textbf{5}, 3049 EP  (2014).

\bibitem{Chekhova:16}
M.~V. Chekhova and Z.~Y. Ou, {\protect\JournalTitle{Adv. Opt. Photon.}}
  \textbf{8}, 104 (2016).

\bibitem{Lemieux:16}
S.~Lemieux, M.~Manceau, P.~R. Sharapova, O.~V. Tikhonova, R.~W. Boyd,
  G.~Leuchs, and M.~V. Chekhova, {\protect\JournalTitle{Phys. Rev. Lett.}}
  \textbf{117}, 183601 (2016).

\bibitem{Linneman:16}
D.~Linnemann, H.~Strobel, W.~Muessel, J.~Schulz, R.~J. Lewis-Swan, K.~V.
  Kheruntsyan, and M.~K. Oberthaler, {\protect\JournalTitle{Phys. Rev. Lett.}}
  \textbf{117}, 013001 (2016).

\bibitem{Lukens:16}
J.~M. Lukens, N.~A. Peters, and R.~C. Pooser, {\protect\JournalTitle{Opt.
  Lett.}} \textbf{41}, 5438 (2016).

\bibitem{Anderson:17}
B.~E. Anderson, P.~Gupta, B.~L. Schmittberger, T.~Horrom, C.~Hermann-Avigliano,
  K.~M. Jones, and P.~D. Lett, {\protect\JournalTitle{Optica}} \textbf{4}, 752
  (2017).

\bibitem{Gong:17}
Q.-K. Gong, X.-L. Hu, D.~Li, C.-H. Yuan, Z.~Y. Ou, and W.~Zhang,
  {\protect\JournalTitle{Phys. Rev. A}} \textbf{96}, 033809 (2017).

\bibitem{Manceau:17PRL}
M.~Manceau, G.~Leuchs, F.~Khalili, and M.~Chekhova,
  {\protect\JournalTitle{Phys. Rev. Lett.}} \textbf{119}, 223604 (2017).

\bibitem{Du:18}
W.~Du, J.~Jia, J.~F. Chen, Z.~Y. Ou, and W.~Zhang, {\protect\JournalTitle{Opt.
  Lett.}} \textbf{43}, 1051 (2018).

\bibitem{Liu:18}
J.~Liu, S.~Li, D.~Wei, H.~Gao, and F.~Li, {\protect\JournalTitle{Journal of
  Optics}} \textbf{20}, 025201 (2018).

\bibitem{Lukens:18}
J.~M. Lukens, R.~C. Pooser, and N.~A. Peters, {\protect\JournalTitle{Applied
  Physics Letters}} \textbf{113}, 091103 (2018).

\bibitem{Liu:18fibers}
Y.~Liu, J.~Li, L.~Cui, N.~Huo, S.~M. Assad, X.~Li, and Z.~Y. Ou,
  {\protect\JournalTitle{Opt. Express}} \textbf{26}, 27705 (2018).

\bibitem{Shaked:18}
Y.~Shaked, Y.~Michael, R.~Z. Vered, L.~Bello, M.~Rosenbluh, and A.~Pe'er,
  {\protect\JournalTitle{Nature Communications}} \textbf{9}, 609 (2018).

\bibitem{Yurke:86}
B.~Yurke, S.~L. McCall, and J.~R. Klauder, {\protect\JournalTitle{Phys. Rev.
  A}} \textbf{33}, 4033 (1986).

\bibitem{Sparaciari:16}
C.~Sparaciari, S.~Olivares, and M.~G.~A. Paris, {\protect\JournalTitle{Phys.
  Rev. A}} \textbf{93}, 023810 (2016).

\bibitem{Gross:10}
C.~Gross, T.~Zibold, E.~Nicklas, J.~Est{\`e}ve, and M.~K. Oberthaler,
  {\protect\JournalTitle{Nature}} \textbf{464}, 1165 (2010).

\bibitem{Chen:15}
B.~Chen, C.~Qiu, S.~Chen, J.~Guo, L.~Q. Chen, Z.~Y. Ou, and W.~Zhang,
  {\protect\JournalTitle{Phys. Rev. Lett.}} \textbf{115}, 043602 (2015).

\bibitem{Manceau:17}
M.~Manceau, F.~Khalili, and M.~Chekhova, {\protect\JournalTitle{New Journal of
  Physics}} \textbf{19}, 013014 (2017).

\bibitem{Giese:17}
E.~Giese, S.~Lemieux, M.~Manceau, R.~Fickler, and R.~W. Boyd,
  {\protect\JournalTitle{Phys. Rev. A}} \textbf{96}, 053863 (2017).

\bibitem{Knyazev:19}
E.~Knyazev, F.~Y. Khalili, and M.~V. Chekhova, {\protect\JournalTitle{Opt.
  Express}} \textbf{27}, 7868 (2019).

\bibitem{Iskhakov:12}
T.~S. Iskhakov, A.~M. P\'{e}rez, K.~Y. Spasibko, M.~V. Chekhova, and G.~Leuchs,
  {\protect\JournalTitle{Opt. Lett.}} \textbf{37}, 1919 (2012).

\bibitem{Beltran:17}
L.~Beltran, G.~Frascella, A.~M. Perez, R.~Fickler, P.~R. Sharapova, M.~Manceau,
  O.~V. Tikhonova, R.~W. Boyd, G.~Leuchs, and M.~V. Chekhova,
  {\protect\JournalTitle{Journal of Optics}} \textbf{19}, 044005 (2017).

\bibitem{Sharapova:15}
P.~Sharapova, A.~M. P\'erez, O.~V. Tikhonova, and M.~V. Chekhova,
  {\protect\JournalTitle{Phys. Rev. A}} \textbf{91}, 043816 (2015).

\bibitem{Zakharov:18}
R.~V. Zakharov and O.~V. Tikhonova, {\protect\JournalTitle{Laser Physics
  Letters}} \textbf{15}, 055205 (2018).

\bibitem{vandeNes:06}
A.~S. van~de Nes, J.~J.~M. Braat, and S.~F. Pereira,
  {\protect\JournalTitle{Reports on Progress in Physics}} \textbf{69}, 2323
  (2006).

\bibitem{Molina:07}
G.~Molina-Terriza, J.~P. Torres, and L.~Torner, {\protect\JournalTitle{Nature
  Physics}} \textbf{3}, 305 EP  (2007).

\bibitem{Boyer:08Scie}
V.~Boyer, A.~M. Marino, R.~C. Pooser, and P.~D. Lett,
  {\protect\JournalTitle{Science}} \textbf{321}, 544 (2008).

\bibitem{Bennink:02}
R.~S. Bennink and R.~W. Boyd, {\protect\JournalTitle{Phys. Rev. A}}
  \textbf{66}, 053815 (2002).

\bibitem{Embrey:15}
C.~S. Embrey, M.~T. Turnbull, P.~G. Petrov, and V.~Boyer,
  {\protect\JournalTitle{Phys. Rev. X}} \textbf{5}, 031004 (2015).

\bibitem{Wasilewski:06}
W.~Wasilewski, A.~I. Lvovsky, K.~Banaszek, and C.~Radzewicz,
  {\protect\JournalTitle{Phys. Rev. A}} \textbf{73}, 063819 (2006).

\bibitem{Sharapova:18}
P.~Sharapova, G.~Frascella, M.~Riabinin, A.~Perez, O.~Tikhonova, S.~Lemieux,
  R.~Boyd, G.~Leuchs, and M.~Chekhova, {\protect\JournalTitle{arXiv preprint
  arXiv:1905.10109}}  (2019).

\bibitem{Armstrong:12}
S.~Armstrong, J.-F. Morizur, J.~Janousek, B.~Hage, N.~Treps, P.~K. Lam, and
  H.-A. Bachor, {\protect\JournalTitle{Nature Communications}} \textbf{3}, 1026
  EP  (2012).

\bibitem{Cai:17}
Y.~Cai, J.~Roslund, G.~Ferrini, F.~Arzani, X.~Xu, C.~Fabre, and N.~Treps,
  {\protect\JournalTitle{Nature Communications}} \textbf{8}, 15645 EP  (2017).

\end{thebibliography}

% Full bibliography will be added automatically on a new page for Optics Letters submissions. This command is ignored for journal article submissions.
% Note that this extra page will not count against page length.
\bibliographyfullrefs{widefieldSu11}

\end{document}